*Review*

# Inorganic Component Imaging of Aggregate Glue Droplets on Spider Orb Webs by TOF–SIMS

Yue Zhao [1,2] , Masato Morita [3] , and Tetsuo Sakamoto [3,*]

[1] *Collaborative Open Research Center, Kogakuin University,*
*2665–1 Nakano, Hachioji, Tokyo, 192–0015 Japan*
[2] *Current address: Laser Science Laboratory, Toyota Technological Institute,*
*2–12–1 Hisakata, Tempaku–ku, Nagoya 468–8511, Japan*
[2] *Department of Applied Physics, School of Advanced Engineering, Kogakuin University,*
*2665–1 Nakano, Hachioji, Tokyo, 192–0015 Japan*
[*] *ct13087@ns.kogakuin.ac.jp*

In this review, we discuss the use of time–of–flight secondary–ion mass spectrometry (TOF–SIMS) technology for analyzing the viscous glue (is called aggregate glue droplets) of spider orb webs and examine the results obtained. Element distribution images of the aggregate glue droplets were observed by TOF–SIMS. A uniform element distribution is seen for suspended pristine aggregate glue droplets, and a differential spreading of aggregate glue components is seen for attached aggregate glue droplets. We also observed TOF–SIMS images of water in aggregate glue droplets, where water was observed to be consistent with the distribution of oozing salt. We also found that the alkali metal in the aggregate glue droplets showed similar characteristics by feeding cesium carbonate to spiders.

## 1. Background

Spiders live in almost every habitat on earth, with the exception of the polar regions and the oceans. So far, at least 48000 spider species have been recorded by taxonomists. The evolution of spiders can be traced back hundreds of millions of years. The most common and widely recognized family of spiders is the family of orb–weaver spiders. It is well known that they capture prey by weaving an "invisible" sticky orb web. The adhesive substance on spider orb webs is a natural composite material with excellent performance. The viscous glue of spider orb webs is considered to be one of the most effective and strongest biological glues [1–4]. For a long time, people have believed that the sticky substance on the orb web is just a simple adhesive. However, recent research has suggested that it may be an "intelligent" material [5]. The viscosity of the glue on the orb web increases dramatically upon the impact of flying prey. There is a functional transformation of the glue from a viscous substance (for capturing prey) to a hardened substance (for retaining prey) [6]. This transformation occurs solely by the stimulation of mechanical forces [7], while the thermodynamic effect is the essence of the transformation. It subverts the perception that the prey is captured only by the stickiness of the glue. It has made people realize the fascination of spiders that have evolved over hundreds of millions of years. This natural composite material has very broad application prospects. It provides a new material for materials science. In addition, an understanding of the adhesion mechanism of the aggregate glue droplets is of great significance for mimicking bioadhesives. Generally, the final chemical state of adhesives determines whether their adhesion process depends on the chemical reaction or physically hardening. For example, the structural adhesives (e.g., epoxy adhesives, acrylic adhesives, and urethane adhesives) cure because of a chemical reaction, and solvent-based adhesives cure because of a solvent or water evaporates. The curing of aggregate glue droplets is more like physically hardening. It is worth mentioning that in wet environments, the aggregated glue droplet possesses a mechanism to remove interfacial water from the contact surfaces in order to form bonds [8]. The ability of the aggregate glue droplets to adhere under wet conditions [4,9–14] provides insight for the design of synthetic adhesives for biomedical applications [15]. Studies of the aggregate glue droplet constituents are still at the level of bulk analysis. However, we have attempted to observe and analyze it by using time–of–flight

secondary–ion mass spectrometry (TOF–SIMS), which is a sensitive surface analysis method. Herein, we discuss interesting findings about the aggregate glue droplets [7,16–18].

## 2. Aggregate Glue Droplets in Orb Webs

After hundreds of millions of years of evolution, spiders have developed ingenious structures on both the macro– and microscales. Orb–weaver spiders are members of the family *Araneidae* [19,20]. As shown in Table 1, the orb–weaver spider typically has seven types of secretory gland, secreting various substances for different functions. Figure 1 illustrates the seven glands, the different silks produced, their functions, and the structure of an orb web. The spider first attaches the dragline to a substrate (concrete, alloy, metal, glass, plant branches, leaves, etc.) by a staple–like attachment disc. The spider will then hang down while secreting dragline from the major ampullate gland, until it lands somewhere determined by the wind. This strong first dragline is the most basic silk for making an orb web. Elongation of the dragline is 20–45% (7.4–16.7 times that of Kevlar and 25–56 times that of steel) and the breaking energy is 111.2–180.9 MJ/m$^3$ (2.2–3.6 times that of Kevlar and 18.5–30.2 times that of steel) [21,22]. The spider then secretes another silk from the center of the base silk to make an overall Y–shaped structure and links the end points of the Y–shaped structure to complete a triangular frame. The spider secretes radial lines and leads them from the center of the "Y" shape to the frame, creating the skeleton for the entire orb web. A "hub" will be made in the center while the radial lines are being woven. The spider then weaves an auxiliary spiral line from the hub to provide a scaffold and a guide line for the next weave. The spacing of the auxiliary spiral line is roughly the length of the spider's body. The frame line, radial line, and auxiliary spiral line created up to this stage are not sticky. Finally, the spider weaves the spiral line along the auxiliary spiral line from outside to inside. While secreting the spiral line, viscous glue will be secreted from the aggregate glands to coat the spiral line. As shown in the inset on the left of Figure 1, the coating of viscous glue is initially uniform, and then it quickly forms equidistant spherical structures in less than tens of seconds as a result of Plateau–Rayleigh instability [4,11,23]. The equidistant spheres of viscous glue are called aggregate glue droplets. Only the spiral lines are sticky in the orb web, and these occupy most of the area. Elongation of spiral line is very high (>200%) [24], 4.5–10 times that of the dragline. Moreover, the breaking energy (131 MJ/m$^3$, without viscous glue; 211 MJ/m$^3$, with viscous glue) [25] of the spiral line is comparable to that of the dragline. Therefore, the spiral line can effectively buffer the impact of high–speed flying prey and can absorb the huge kinetic energy introduced by the prey [24,26]. The spiral line is restored to its initial shape in a short time after being impacted because of its excellent toughness [24]. However long it is stretched, the silk recoils back when it relaxes, with no sagging caused by being too soft and no curving as a result of being too hard [24]. Under extension, the spiral line behaves as a stretchy solid, but it totally switches behavior self–adapts in compression to now become liquid–like [27]. Each of the aggregate glue droplets can act like a "windlass" to spool and pack the spiral line silk [24,27]. Thus, the silk and the whole web can remain under tension. Orb–weaver spiders employ clever behavioral strategies by coating a viscous glue onto the almost invisible orb web to capture prey. When the unlucky prey fly into the orb web, they get stuck and cannot break free; this increases the foraging success of the spider [28]. In addition, spiders use their orb webs very economically, and orb webs are ingested by the spider for recycling [29,30]. This behavior is common in some orb–weaver spiders.

**Table 1.** Glands, secretion components, and secretion functions in orb–weaver spiders

| Gland | Secretion components | Functions |
|---|---|---|
| Major ampullate | MaSp (protein) | Dragline, frame line, radial line |
| Minor ampullate | MiSp (protein) | Auxiliary spiral line (guide line) |
| Flagelliform | Flag (protein) | Capture spiral line |
| Aggregate | ASG (glycoprotein), LMMCs*, water | Sticky substance |
| Piriform | PySp (protein), glycoproteins, lipids | Attachment cement |
| Aciniform | AcSp (protein) | Prey wrapping, inner–layer silk of egg sac |
| Cylindrical | CySp (protein) | Outer–layer silk of egg sac |

\* low–molecular–mass compounds

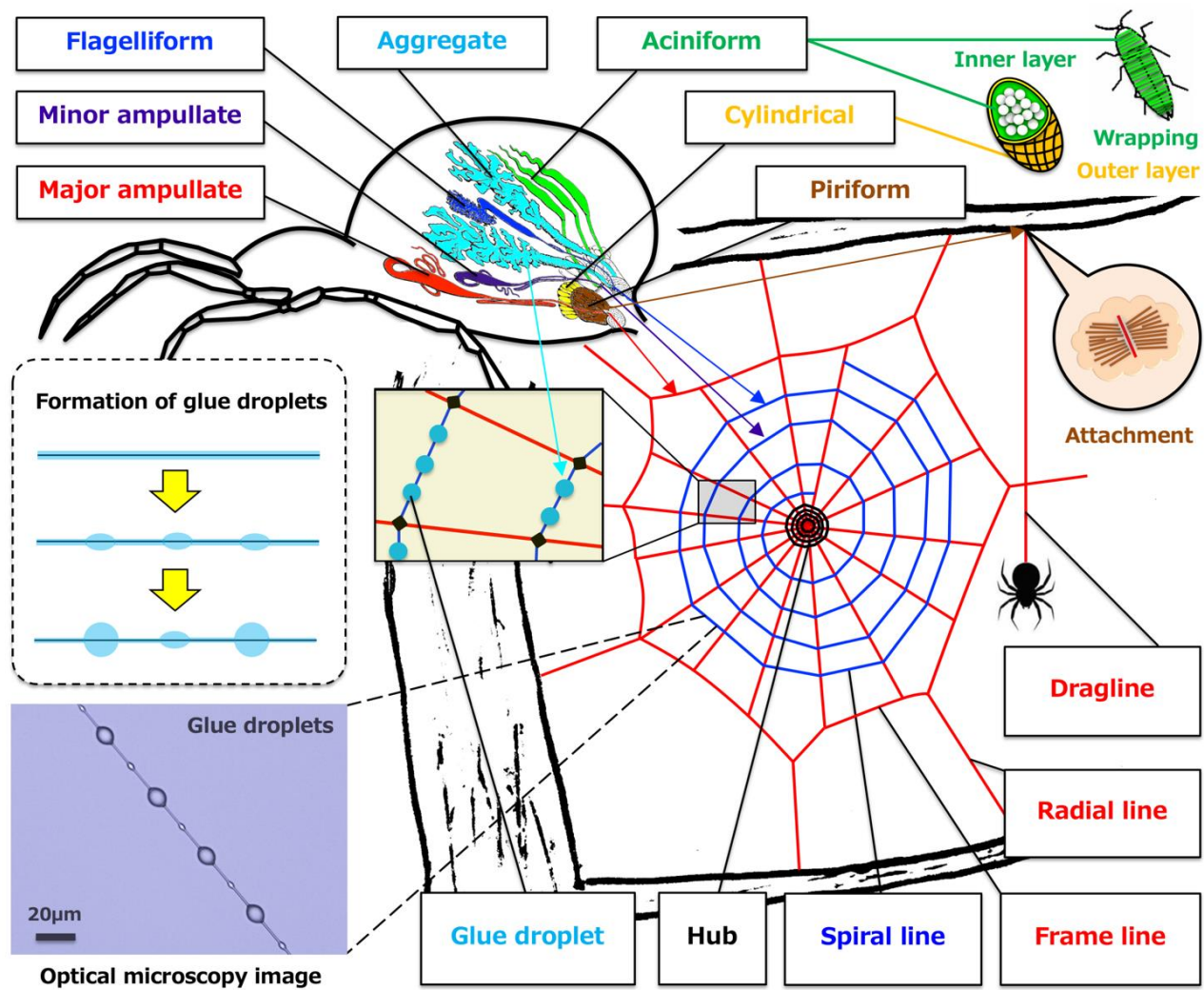

**Figure 1.** Illustration of a typical orb–weaver spider's seven types of glands, the different silks produced, and their functions. The dragline and structural silk (frame lines, radial lines) are secreted from the major ampullate gland. The auxiliary spiral line is secreted from the minor ampullate gland. Attachment cement is secreted from the piriform gland for anchoring the dragline and structural silk. The capture spiral line is secreted from the flagelliform gland and is coated with sticky substance that is secreted from the aggregate gland. The sticky substance will form equidistant glue droplets on the spiral line as a result of Plateau–Rayleigh instability. The egg sac is wrapped with a tough outer–layer silk, which is secreted from the cylindrical gland. Aciniform silk is used in prey wrapping and for the inner layer of the egg sac. The center of the orb web is called the "hub". The spider usually stands on the hub. (Color version online.)

The aggregate glue droplet structure of spider orb webs can be traced back to the early Cretaceous Hauterivian age (approximately 130 million years ago) [31,32]. After more than 100 million years of evolution, the aggregate glue droplets have evolved into a natural material with outstanding properties [6]. Although spiders inhabit a variety of environments, including mountains, grasslands, rainforests, wetlands, coasts, and cities, their aggregate glue droplets can effectively attach to and capture the prey in all of them. The mechanical properties of the aggregate glue droplets play an important role in adhesion [6]. The viscosity of the aggregate glue droplets increases dramatically upon the impact of flying prey [6]. The glue then solidifies and prevents the prey from escaping [6]. This suggests that the glue is most likely a non–Newtonian fluid [18]. However, the properties of aggregate glue droplets are not fully understood. Bulk analyses of aggregate glue droplets show that the main components are two glycoproteins (ASG–1 and ASG–2) [2,28,33–36], low–molecular–mass compounds (LMMCs) (GABamide [$\gamma$–aminobutyramide], *N*–acetyltaurine, choline, betaine, isethionic acid, and pyrrolidone) [3,13,37–40], inorganic salts (potassium nitrate and potassium dihydrogen phosphate) [3,38,39], and water [3,13]. The glycoproteins are *O*–glycosylated and have an *N*–acetylgalactosamine residue *O*–linked to a threonine residue [2]. The two glycoproteins (ASG–1 and ASG–2) are expressed from opposite strands of the same DNA sequence [34]. In the nonrepetitive region, ASG–1 has a high proportion of charged amino acids, and the upstream nonrepetitive sequence has chitin binding traits [34]. The ASG2 repetitive domain is similar to the capture spiral line silk protein repetitive sequence domains [34]. The high proportion of proline in ASG2 and the similarity between its *N*–terminal region sequence and that of collagen suggest that it likely possesses elasticity [34]. Generally, the LMMCs and inorganic salts are collectively called "salts". Salts play an important role in the versatility of the aggregate glue droplet. Denaturation of glycoproteins caused by loss of salts has been confirmed [41,42]. The shape and characteristics of aggregate glue droplets will change greatly under mechanical stimulation. Even if it is simply attached to the substrate, an aggregate glue droplet will stick and spread after contact with the substrate over a period of microseconds to seconds. The microscopy images in Figure 2 show the appearance of aggregate glue droplets before and after adhesion to substrates. Hygroscopic LMMCs in the glue droplet can absorb atmospheric water and solvate the glycoproteins, causing them to spread and adhere to prey [13]. These component analysis results are based on bulk analysis (nuclear magnetic resonance (NMR) spectroscopy, etc.). However, bulk analysis destroys the location information, so distribution information for the various components cannot be obtained. Therefore, the analysis of aggregate glue droplets is incomplete; in particular, the distribution of inorganic matter has not been fully analyzed by bulk analysis. TOF–SIMS can be used to analyze the inorganic matter in aggregate glue droplets.

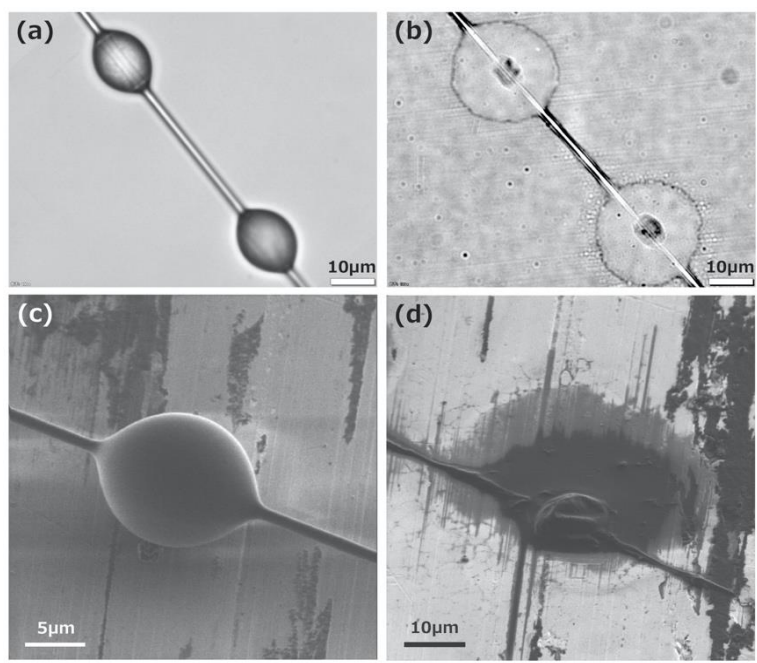

**Figure 2.** Optical microscopy images and secondary electron images of aggregate glue droplets. (a) Optical microscopy image of suspended pristine aggregate glue droplets. (b) Optical microscopy image of aggregate glue droplets adhered to a glass substrate. (c) Secondary electron image of a suspended pristine aggregate glue droplet. (Re–printed Figure S5 of the supporting information from ref. [7] with permission. Copyright 2019 by the Japan Society for Analytical Chemistry.) (d) Secondary electron image of an aggregate glue droplet adhered to an aluminum substrate. (Re–printed Figure S5 of the supporting information from ref. [7] with permission. Copyright 2019 by the Japan Society for Analytical Chemistry.)

## 3.  Samples and Focused-Ion-Beam (FIB)-TOF-SIMS

### 3.1. Sample Collection and Handling

In Japan, orb–weaver spiders usually appear between April and November. Figure 3(a) and (b) shows photographs of a spider of the *Araneus* genus and its orb web. *Araneus* is a genus of common orb–weaver spiders that are distributed all over Japan. In cities, they can be found in parks and on guardrails under the street lights of bridges and roads. *Araneus* spiders mature in summer, so it is easy to find their orb webs then. Generally, they will start weaving orb webs at night and recycle the webs before dawn. Therefore, fresh orb webs can be collected each night. Spiders usually stay in the hub at the center of the orb web. When we collected an orb web, we drove the spider away from the web, placed a ring on the orb web, and then cut off the peripheral silk. The orb web is left on the ring because of its own stickiness (see Figure 3(c)). The ring can be metal or other hard materials. Samples should be kept at the humidity and temperature of the external environment during transportation back to the laboratory. If the collection site is not very far from the laboratory, the sample can be exposed to the external environment during transportation. Because aggregate glue droplets are very sensitive to humidity, they should be measured or frozen immediately upon arrival at the laboratory. The sample needs to be installed on a dedicated holder for TOF–SIMS analysis. As shown in Figure 3(d), the spiral line is first adhered to the edge of the holder, and then the height of the aluminum pellet is adjusted with screws. A part of the spiral line will first adhere to the aluminum pellet, and some aggregate glue droplets remain suspended because of the height difference. The aggregate glue droplets are electrically conductive, and necks of liquid connect the droplets on the spiral line [43]. Hence, suspended glue droplets will not cause charge up. In order to maintain the humidity and temperature, sample handling can be performed outdoors.

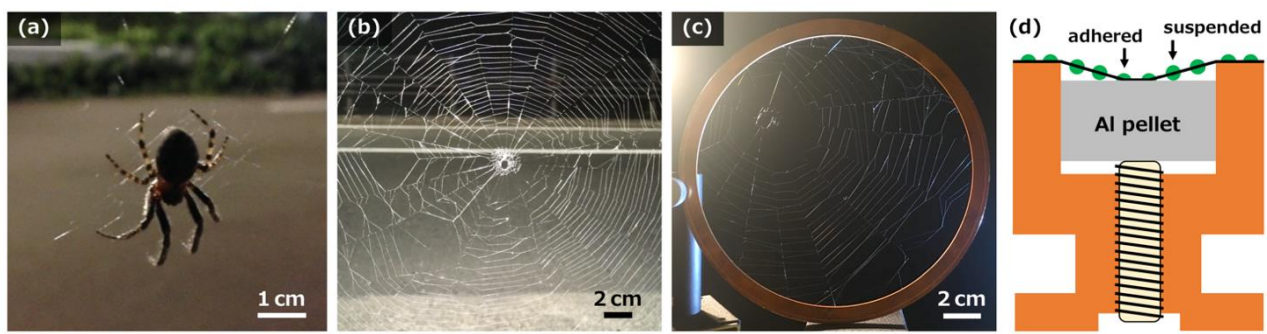

**Figure 3.** (a) Photograph of a spider (*Araneus* genus). (Re–printed Figure 1 from ref. [7] with permission. Copyright 2019 by the Japan Society for Analytical Chemistry.) (b) Photograph of the orb web of a spider (*Araneus* genus). (Re–printed Figure 1 from ref. [7] with permission. Copyright 2019 by the Japan Society for Analytical Chemistry.) (c) Photograph of the collected orb web. (Re–printed Figure 1 from ref. [7] with permission. Copyright 2019 by the Japan Society for Analytical Chemistry.) (d) Schematic illustration of sample handling. (Color version online.)

The composition and adhesion characteristics of aggregate glue droplets from spiders in the genera *Larinioides* [6,8–10,38,41,44,45], *Araneus* [2,4,24,33, 34,38,46], *Nephila* [34], *Argiope* [3,11,14,36,37,45–49],

*Neoscona* [39], and *Araneoidea* [50,51] have been reported. The spiders above and the *Araneus* spider introduced in this review all belong to the family *Araneidae* [28,50]. It should be noted that, although the adhesion mechanism of aggregate glue droplets is similar among different species, differences in composition between the species cannot be ruled out.

### 3.2. FIB–TOF–SIMS

TOF–SIMS is an excellent surface analysis method. In particular, TOF–SIMS with a focused Ga ion beam is effective for analyzing small particles [52–70]. An FIB–TOF–SIMS instrument developed by Sakamoto *et al.* was used for component analysis and elemental mapping of aggregate glue droplets on orb webs [7,16–18]. Figure 4 shows a schematic illustration of the FIB–TOF–SIMS instrument. The ion source for the FIB is $Ga^+$, the acceleration voltage is 30 keV, and the beam angle is 45˚. Generally, we first look for the target through real–time scanning ion microscope (SIM) image with a large field of view (500 μm × 500 μm). After locking the target, we zoom in to the proper field of view. Because excessive beam current will cause the sticky ball to deform or will cause damage, the beam current (DC) should be kept to <0.2 nA and the target should be locked in the shortest time possible. In TOF–SIMS analyses with a pulse mode, the beam current (DC) can be set between 0.1 and 0.7 nA. Normally, during mapping, the FIB pulse–time width is set to 200 ns and the frequency is set to 10 kHz. Delayed extraction is adopted. The secondary ion extractor is 8 mm above the sputtering site of the sample. A TOF–MS ion extraction voltage (–1.8 kV) is applied and maintained for 5 µs. The voltage of the flight tube is –2.5 kV, and the reflection potential is –0.86 kV. A time–to–digital (TDC) converter is used to process the signal detected by the microchannel plate (MCP). Elemental mapping is performed by scanning the sample surface with the FIB. In elemental mapping, the best lateral resolution is about 40 nm. The mass resolution is $m/\Delta m$ = 4000 (at $m/z$ 28).

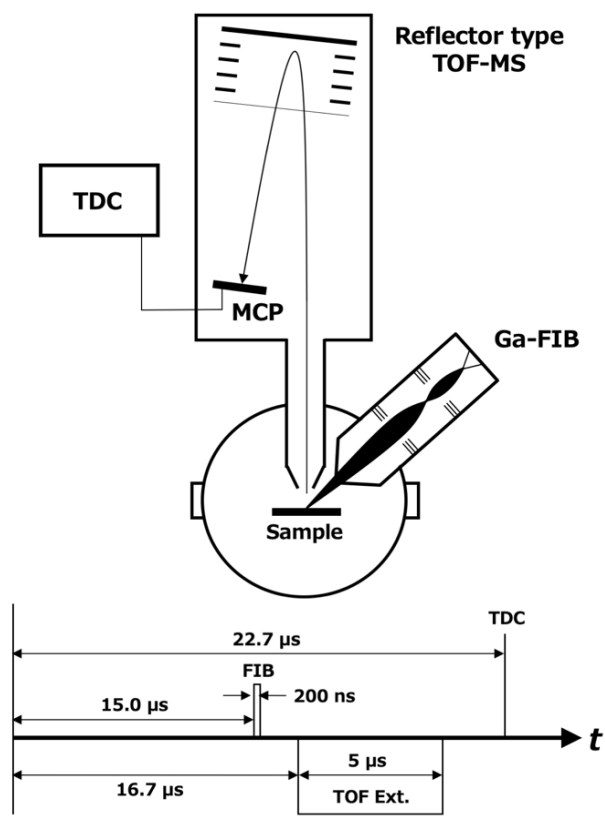

**Figure 4.** Schematic illustration of the FIB–TOF–SIMS instrument. Time settings are shown at the bottom. FIB: focused–ion beam; MCP: microchannel plate; TDC: time–to–digital converter; TOF–MS: time–of–flight mass spectrometry.

### 3.3. Freezing Method for TOF–SIMS Analysis

Normally, the vacuum pressure of the TOF–SIMS analytical chamber is $10^{-6}$ Pa. Because the aggregate glue droplets contain water, it is necessary to lower the temperature of the sample below the sublimation point of water at the pressure of $10^{-6}$ Pa. Figure 5 shows a flash freezing method for TOF–SIMS. The stage can be cooled to –130°C by circulating liquid nitrogen. The holder with the handled sample and cap are immersed into liquid nitrogen. After sufficient cooling, the cap is screwed to the holder within liquid nitrogen. The holder is then quickly introduced into the load–lock chamber. The load–lock chamber is evacuated to vacuum. Because of the high pressure caused by evaporation of liquid nitrogen in the cap, the sample is not contaminated with outside air. When the vacuum in the load–lock chamber reaches $10^{-4}$ Pa, the cap is unscrewed, and the holder is then introduced into the analytical chamber. For free water, amorphous ice is formed in the case of flash freezing [71], whereas ice crystals are formed in the case of slow freezing. When performing the slow freezing, the holder with the handled sample is put into the freezer compartment of a refrigerator first. After a few hours of freezing, the procedure for flash freezing described above is executed. The aggregate glue droplets will not deform under vacuum, even without freezing, so they can also be analyzed by TOF–SIMS at room temperature; however, samples should be frozen for characterization of water.

### 3.4. FIB Micromachining

In the TOF–SIMS analysis of aggregate glue droplets, it is necessary to analyze not only the surface but also the interior while maintaining the aggregate glue droplet shape. The inside of an aggregate glue droplet can be exposed by cutting with FIB micromachining. Figure 6 shows FIB micromachining of an aggregate glue droplet after flash freezing and at room temperature (21°C). The cross section of the aggregate glue droplet is very flat after cutting, which is very important. An uneven surface will cause an artefact in TOF–SIMS mapping, with the contrast at the edges and the raised portions appearing relatively strong. The aggregate glue droplets are very easy to cut by FIB micromachining, and a large beam current is not required. In Figure 6, the beam current for roughing was 0.64 nA, and the beam current for finishing was 0.11 nA.

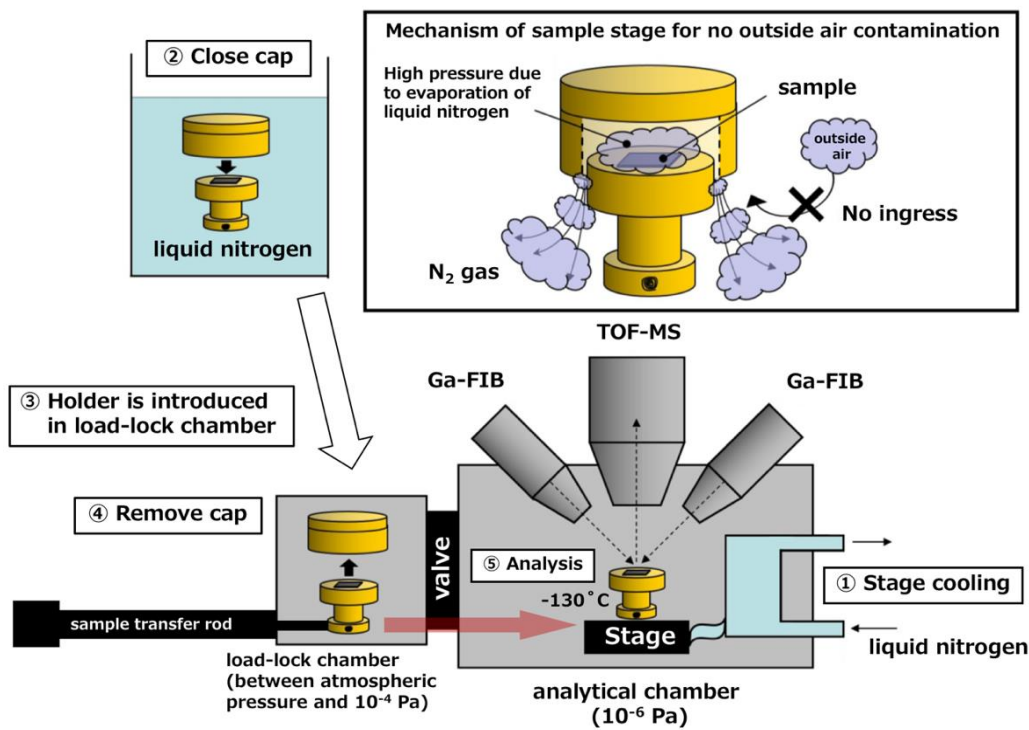

**Figure 5.** Procedure for flash freezing. FIB: focused–ion beam; TOF–MS: time–of–flight mass spectrometry. (Re–printed Figure S2 of the supporting information from ref. [7] with permission. Copyright 2019 by the Japan Society for Analytical Chemistry.) (Color version online.)

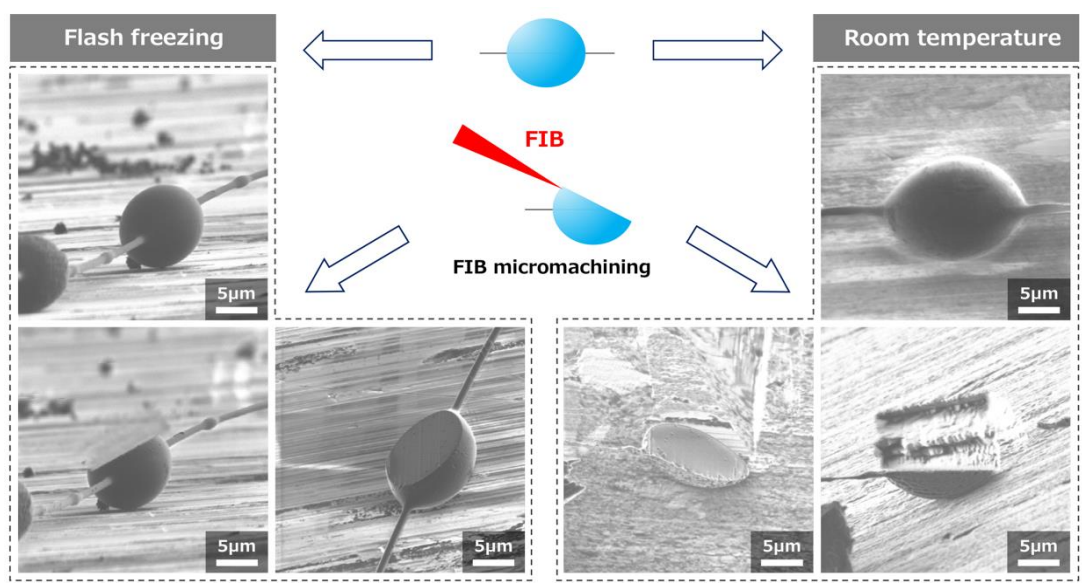

**Figure 6.** Secondary electron images of aggregate glue droplets before and after FIB micromachining after flash freezing and at room temperature (21°C). FIB: Focused–ion beam. (Adapted Figure S4 of the supporting information from ref. [7] with permission. Copyright 2019 by the Japan Society for Analytical Chemistry.) (Color version online.)

## 4. TOF–SIMS on Aggregate Glue Droplets

### 4.1. Suspended Pristine Aggregate Glue Droplets

As is well known, under room temperature conditions, the water in the vacuum chamber ($10^{-6}$ Pa) of TOF-SIMS equipment evaporates rapidly. In the case where it is not clear whether the water in the pristine aggregate glue droplet is bound water or free water, it is necessary to conduct the experiment under freezing conditions. As a result, the water signal was not detected in suspended pristine aggregate glue droplets under freezing conditions. A control experiment was conducted at room temperature. Figure 7 shows TOF–SIMS images of the inorganic salts in suspended pristine aggregate glue droplets. Na$^+$ ($m/z$ 23), K$^+$ ($m/z$ 39 and 41), Cl$^-$ ($m/z$ 35 and 37), and CN$^-$ ($m/z$ 26) were detected from the surface (see Figure 7 (a)) and on the inside (cross section) (see Figure 7 (b)) under flash freezing. In the TOF–SIMS results, the CN$^-$ ($m/z$ 26) was assigned to protein. After freezing, Na$^+$ and K$^+$ were the only inorganic cations observed. This is consistent with the previous results [72] of a SIMS analysis of the flagelliform protein. By contrast, at room temperature (21°C), in addition to Na$^+$ ($m/z$ 23), K$^+$ ($m/z$ 39 and 41), Cl$^-$ ($m/z$ 35 and 37), CN$^-$ ($m/z$ 26) and Ca$^+$ ($m/z$ 40) was detected. The Ca$^+$ signal is only detected at room temperature (21°C) (Figure 7 (c) and (d)). Regardless of whether the droplet was flash frozen or is at room temperature (21°C), the distributions of Na$^+$, K$^+$, the counter ion Cl⁻, and the proteins are uniform from the surface to the inside of the droplet. However, the distribution of Ca⁺ is significantly different. At room temperature (21°C), Ca⁺ exhibits nanoparticle distribution at the surface and inside the aggregate glue droplets (Figure 7 (c) and (d)). This suggests that calcium may be a key element for low–temperature sensing in the aggregate glue droplets [16]. On the other hand, the spiral lines attached to the aggregate glue droplets remained flexible and were hardly breakable even when immersed in liquid nitrogen (–169°C) [7]. This indicates a low content of free water and suggests that the aggregate glue droplets will not form brittle ice crystals and/or amorphous ice, even in a low–temperature environment far below freezing. Under low–temperature stress, reorganization of calcium may occur, which makes it difficult to detect in the form of Ca⁺ secondary ions. Calcium present in the aggregate glue droplets as nanoparticles may play a key role in preventing crystallization of the aggregate glue droplets [16]. In addition, although the aggregate glue droplets contain $KNO_3$ and $KH_2PO_4$ [3,38,39], the peaks for nitrate ($NO^-$ ($m/z$ 30) and $NO_2^-$ ($m/z$ 46)) and phosphate ($PO^-$ ($m/z$ 47), $PO_2^-$ ($m/z$ 63), and $PO_3^-$ ($m/z$ 79)) were not detected in the suspended pristine aggregate glue droplets.

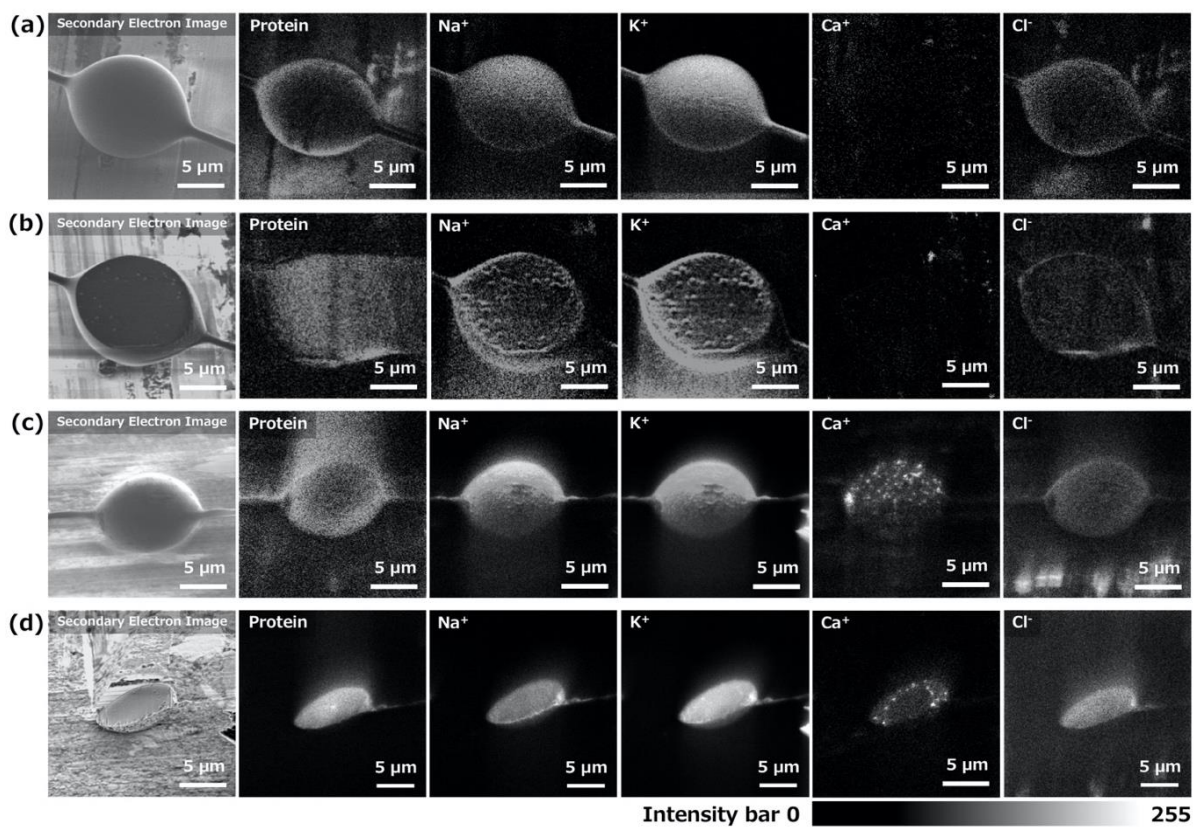

**Figure 7.** Secondary electron images and mapping information for the elements in aggregate glue droplets. (a) Suspended pristine aggregate glue droplet after flash freezing. (b) Cross section of the aggregate glue droplet after flash freezing. (c) Suspended pristine aggregate glue droplet at room temperature (21°C). (d) Cross section of the aggregate glue droplet at room temperature (21°C). (Adapted Figure 2 from ref. [7] with permission. Copyright 2019 by the Japan Society for Analytical Chemistry, and adapted Figure 1 from ref. [16] with permission. Copyright 2019 by the Surface Analysis Society of Japan.)

## 4.2. Attached Aggregate Glue Droplets

The adhesion process of the aggregate glue droplet is completed within a few microseconds. However, the time required to collect one mapping data with TOF-SIMS is far more than a few microseconds. Therefore, the current TOF-SIMS technology is still difficult to capture the dynamic process of aggregate glue droplet adhesion in such a short time. Here, we compare the suspended pristine aggregate glue droplet and the state after the adhesion process. The shape of the aggregate glue droplet changed significantly after attachment to a substrate (see Figure 2). Figure 8 shows the element distributions in an aggregate glue droplet attached to an aluminum substrate at room temperature (21°C). In comparison with the results for the suspended pristine aggregate glue droplet at room temperature (21°C) (see Figure 7(c)), not only were Na⁺ ($m/z$ 23), K⁺ ($m/z$ 39 and 41), Ca⁺ ($m/z$ 40), Cl⁻ ($m/z$ 35 and 37), and CN⁻ ($m/z$ 26) detected, but also P⁻ ($m/z$ 31) and phosphate ($PO^-$ ($m/z$ 47), $PO_2^-$ ($m/z$ 63), and $PO_3^-$ ($m/z$ 79)). Although there are reports [3,38,39] claiming that $KNO_3$ is also present in the glue droplets, the assignable peaks for nitrate ($NO^-$ ($m/z$ 30) and $NO_2^-$ ($m/z$ 46)) were not detected in TOF–SIMS results for aggregate glue droplets from *Araneus* spiders. This may be related to differences between species.

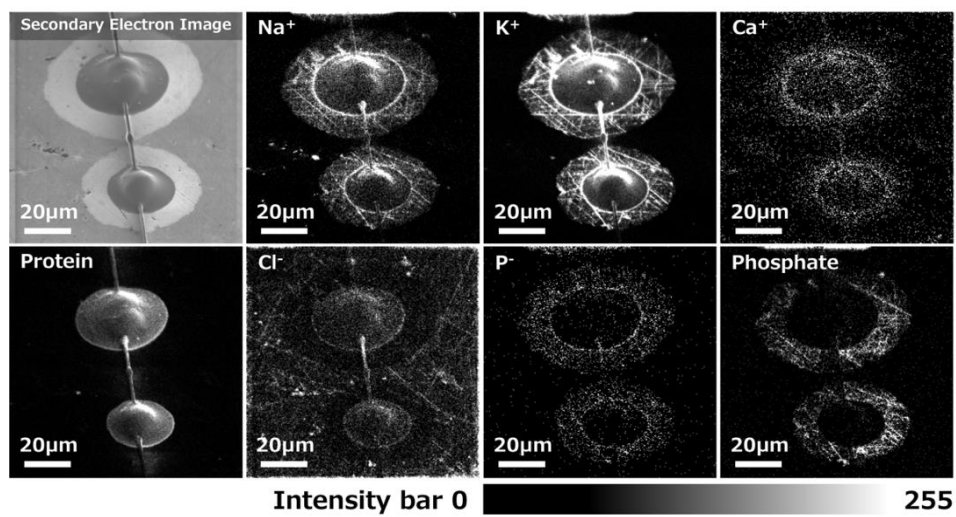

**Figure 8.** Secondary electron image and mapping information for the contained elements in an aggregate glue droplet attached to a substrate at room temperature (21°C)

After adhering to the substrate, the aggregate glue droplet divides into several regions, forming a "fried egg" structure: a center ("yolk") composed of proteins connected to the silk and an outer ring ("egg white") composed of phosphate. Chlorine spreads over a wider region outside the outer ring. The results in Figure 8 suggest phase separation of the inorganic salts and glycoproteins [7]. In this case, the phase separation of the inorganic salts and glycoproteins is caused by mechanical stimulation [7], and can be explained by thermodynamic effects. The spreading speed of water-soluble components (LMMCs) is faster than that of glycoprotein, which results in a differential spreading of aggregate glue components [73]. This spreading effect would be increased on hydrophilic substrates [73]. Besides, the greater the humidity, the larger the area in which the aggregate glue droplet spreads along the substrate [4]. After contact with the object, the LMMCs of the aggregate glue droplet spread and accumulate on the object [42]. Loss of LMMCs causes the glycoproteins to collapse, harden, and lose adhesion [42]. The hardened glycoproteins formed the "yolk" part of the "fried egg" structure in Figure 8. The LMMCs and proteins are likely segregated in the adhered aggregate glue droplet [4,50]. Therefore, the LMMC distribution is likely to be consistent with the phosphate in the outer ring. This can be verified from the water distribution in adhered aggregate glue droplets [18], as described in the following section. In summary, before the prey hits the web, the function of the aggregate glue droplet is stickiness. After contact, phase separation leads to the loss of salts, causing denaturation of the glycoprotein. The function of the aggregate glue droplet changes from stickiness to fixed prey. Furthermore, aggregate glue droplets with low viscosity are spread more readily due to capillary forces [73]. In Figure 8, some bright line–like distributions of sodium, potassium, chlorine, and phosphate can be observed in the outer ring. These lines are scratches on the aluminum substrate. The width of these scratches is on the order of nanometers; hence, it can be concluded that the fluidity of the salts is extremely good. The excellent fluidity allows the aggregate glue droplets to maximize the surface area for adhesion to effectively cope with complex shapes of prey.

### 4.3. Water in Aggregate Glue Droplets

The aggregate glue droplets contain a lot of water [3,4,36,44,46,50,74,75]. NMR data show water peaks in the aggregate glue droplets [42]. Expansion and contraction of suspended aggregate glue droplets are dependent on atmospheric humidity [10–14]. These observations suggest that the aggregate glue droplets are hygroscopic. However, the characterization of water in aggregate glue droplets, especially the distribution, is very challenging. Once an aggregate glue droplet is touched, its shape will be destroyed. Hygroscopic LMMCs will be lost, and the properties of the entire aggregate glue droplet will change. However, the water can be characterized by freezing in TOF–SIMS analysis [16]. Here the flash and slow freezing method were applied. In TOF–SIMS analysis, water clusters are generally detected by the peaks of $H_2O^+$ ($m/z$ 18) and $H_3O^+$ ($m/z$ 19) [76–82]. In the case of binary salts, the signals of $K^+$ ($m/z$ 39) and $K(KCl)^+$ ($m/z$ 113) are very intense [83]. Because of the presence of $K^+$ and $Cl^-$ in the aggregate glue droplets, KCl and/or KCl·$H_2O$ precipitate under slow freezing and be detected as $K(KCl)^+$ ($m/z$ 113). Unprecipitated $K^+$ ($m/z$ 39) will be combined with free water and detected as $K(H_2O)^+$ ($m/z$ 57) [16]. The distribution of $K(KCl)^+$ and $K(H_2O)^+$ will complement each other [16]. By contrast, under flash freezing, KCl and/or KCl·$2H_2O$ will have no time to precipitate, so only $K(H_2O)^+$ will be detected [18]. The distribution of water in an aggregate glue droplet can be indirectly judged by a comparison of the distribution of $K(KCl)^+$ ($m/z$ 113) and $K(H_2O)^+$ ($m/z$ 57). In a suspended pristine aggregate glue droplet, $K(KCl)^+$ and $K(H_2O)^+$ peaks were not detected under flash freezing (Figure 9 (a) and (b)) or under slow freezing (Figure 9 (c) and (d)). This suggests that the water in the pristine aggregate glue droplet is unlikely to be free water [18]. In addition, liquid–like water peaks were not observed by sum–frequency vibrational spectroscopy, even when pristine aggregate glue droplets were humidified with $H_2O$ vapor [8]. Because sum–frequency vibrational spectroscopy is very sensitive to water interfaces [84], we can conclude that the free water content in the pristine aggregate glue droplets is very low. This low content of free water in pristine aggregate glue droplets provides a basis for the phenomenon that these droplets remain flexible and hardly breakable in liquid nitrogen. By contrast, after adhesion of a droplet to the substrate, distributions of $K(KCl)^+$ and $K(H_2O)^+$ were detected in the outer ring region (corresponding to the distribution of $K^+$). After flash freezing, the distribution of $K(H_2O)^+$ in the outer ring region is very obvious (Figure 9 (e)), but $K(KCl)^+$ was not detected. With slow freezing, although the signals of $K(H_2O)^+$ and $K(KCl)^+$ are weak (Figure 9 (f)), the distribution can still be observed. However, LMMCs, especially *N*–acetyltaurine, choline, and betaine, are the main hygroscopic components [3,4,40], so the distribution of $K(H_2O)^+$ can correspond to the distribution of LMMCs. In this case, the water in $K(H_2O)^+$ must be free water. In the TOF–SIMS results, the phenomenon of water in the aggregate glue droplets is similar to that of phosphate; neither of them can be detected in the suspended pristine aggregate glue droplet, but only after adhesion. The above results suggest that the water in the pristine aggregate glue droplets is more likely to be bound water than free water. The matrix effect caused by bound water makes it difficult to detect water clusters in the pristine aggregate droplets by TOF–SIMS [16].

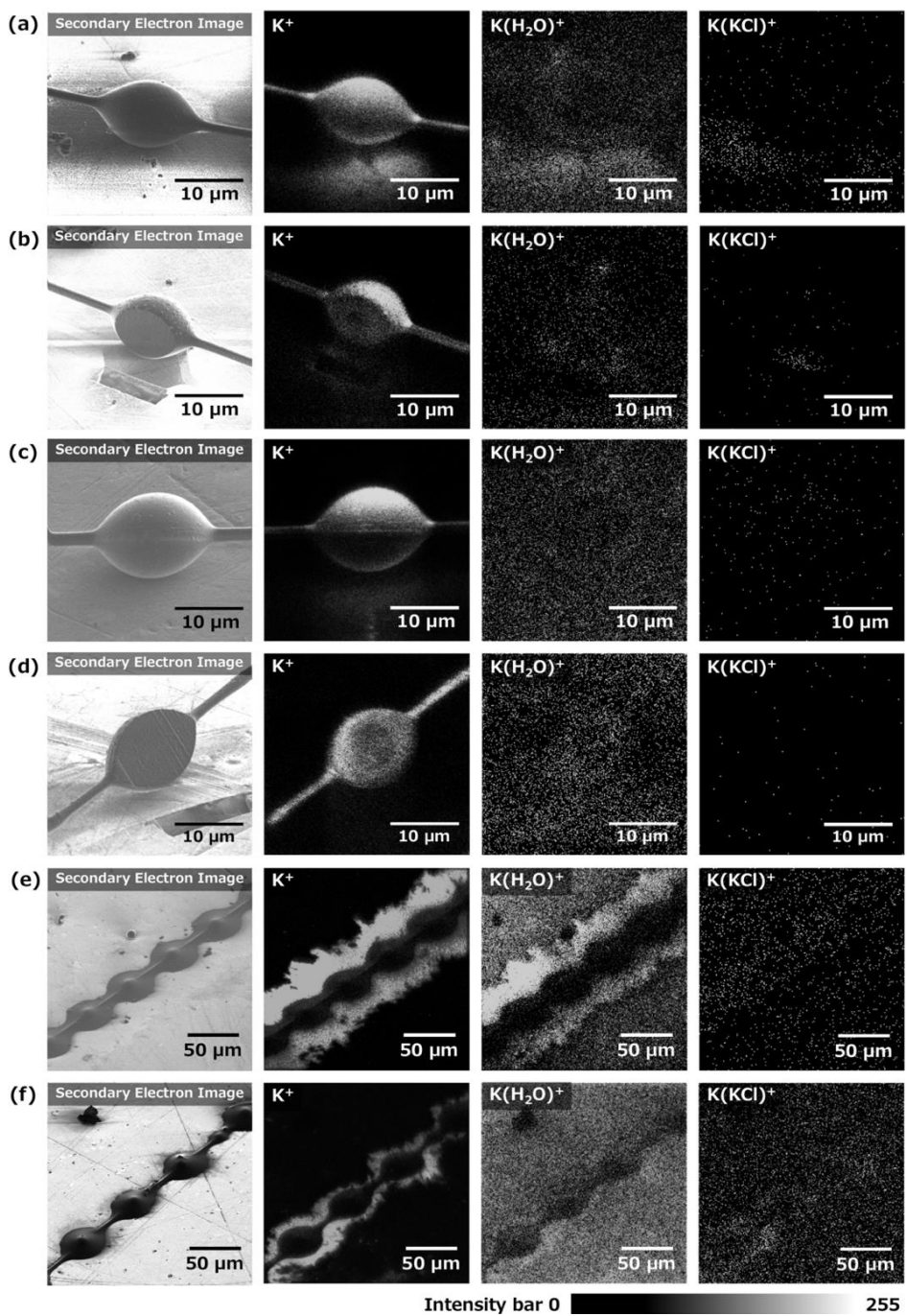

**Figure 9.** Secondary electron images and mapping information for the elements in aggregate glue droplets. (a) Suspended pristine aggregate glue droplet after flash freezing. (b) Cross section of the aggregate glue droplet after flash freezing. (c) Suspended pristine aggregate glue droplet after slow freezing. (d) Cross section of the aggregate glue droplet after slow freezing. (e) Aggregate glue droplet attached to the substrate after flash freezing. (f) Aggregate glue droplet attached to the substrate after slow freezing. (Re–printed Figure 2 and 3 from ref. [18] with permission. Copyright 2020 by John Wiley & Sons, Ltd.)

In addition, a previous report claimed that very weak water clusters were detected at the surface of a pristine aggregate glue droplet [7]. In later research, we found that surface water clusters can be caused by the sample being put into the liquid nitrogen too slowly during the flash freezing operation, so water vapor condensed near the liquid nitrogen and was attached to the surface of the pristine aggregate glue droplet. If the sample is put into the liquid nitrogen fast enough, condensed water vapor can be prevented from attaching to the surface of the pristine aggregate glue droplet.

However, even in humid habitats, aggregate glue droplets can still overcome the interfacial failure of the interfacial bound water to provide adhesion to capture prey [8,40]. In this situation, it should be noted that the layer of free water wrapped around the suspended pristine aggregate glue droplet may be lost during freezing. However, the free water layer may be retained by performing flash freezing in a high–humidity environment. In addition, even if the aggregate glue droplet is exposed to conditions where the relative humidity exceeds 90%, a "fresh" aggregate glue droplet will lose water [85]. The volume of a "fresh" aggregate glue droplet will be reduced by 90% in 30–60 minutes, and this loss is irreversible [85]. Spiders weave an orb web, use it for at least one night, and recycle it before dawn. Thus, "fresh" aggregate glue droplets are not the "normal" state in use; the "normal" state of the droplets for capturing prey is reached after volume reduction.

### 4.4. Compatibility of Alkali Metal Elements in Aggregate Glue Droplets

Spiders weave new orb webs by recycling used ones. Thus, the inorganic salts (sodium and potassium from the alkali metals group) are also recovered during the recycling. If pollutants are adhered to the orb web, then the spiders themselves will become accumulators of such pollutants [43]. Cesium was detected in the aggregate glue droplet of an orb web after the spider had been fed with cesium carbonate [17]. The absorbed cesium was detected on the surface (Figure 10 (a)) and inside (cross section) (Figure 10 (b)) of the aggregate glue droplet; it showed similar behavior to sodium and potassium and was evenly distributed in the aggregate glue droplet. Therefore, *Araneus* spiders can ingest and accumulate cesium, and they can integrate it into their viscous glue. Moreover, the presence of cesium does not change the adhesive properties of the aggregate glue droplets [17]. We conclude that the alkali metal elements in the aggregate glue droplet are compatible.

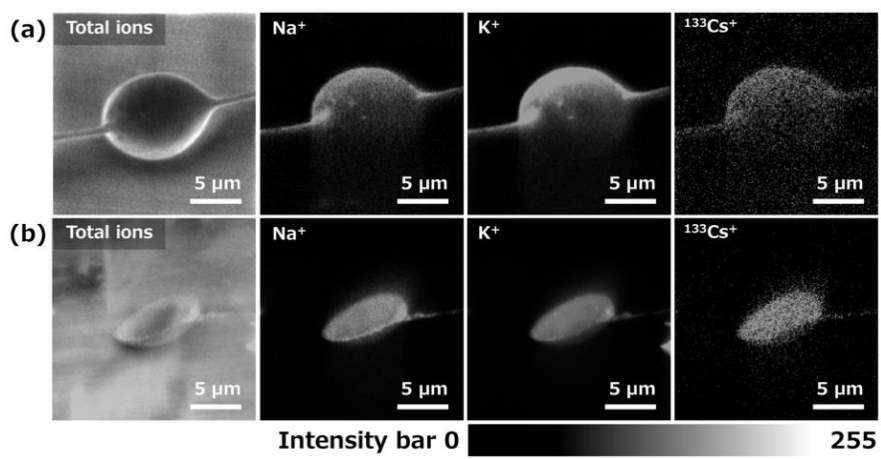

**Figure 10.** (a) Mapping of elements on a suspended pristine aggregate glue droplet surface after flash freezing. (b) Mapping of elements for a cross section of a suspended pristine aggregate glue droplet after flash freezing. (Adapted Figure 2 and 3 from ref. [17] with permission. Copyright 2020 by John Wiley & Sons, Ltd.)

## 5.  Summary and Future Expectations

Research on aggregate glue droplets has changed people's view that the adhesive of spider orb webs is only a simple sticky substance similar to an adhesive tape. The fascination of spiders that have evolved over hundreds of millions of years needs to be reassessed. In this review, we have summarized the most recent studies for the observation of the inorganic salt distribution in the aggregate glue droplets of spider orb webs by using TOF–SIMS [7,16–18], and we have discussed the specific methods of observing aggregate glue droplets with TOF–SIMS. In suspended pristine aggregate glue droplets, the inorganic salts are evenly distributed and the water is unlikely to be free. On the other hand, after adhesion, phase separation of the salts and glycoproteins occurred. The salt exudate absorbed environmental water. An understanding of the adhesion mechanism of aggregate glue droplets is of great significance for mimicking bioadhesives. In particular, the ability of the glue to adhere under wet conditions could provide insight for the design of synthetic adhesives for biomedical applications. TOF–SIMS provides a new method for analyzing aggregate glue droplets. In subsequent experiments, laser sputtered neutral mass spectrometry (SNMS) can be used to analyze the organic matter in aggregate glue droplets. In addition, in view of the spiders' behavior of recycling their used orb webs, heavy water ($D_2O$) can be fed to spiders. From a comparison of the peaks of $K(H_2O)^+$ (*m/z* 57) and $K(D_2O)^+$ (*m/z* 59), it can be determined whether the water is suspended water from the pristine glue drop itself or absorbed atmospheric water after adhesion. Finally, the accumulation of cesium confirmed the feasibility of spider orb webs as an environmental monitor. This may provide a new monitoring method for measuring the radioactive cesium from nuclear accidents in the environment.

## 6.  References

[1]  H. M. Peters, Naturwissenschaften **82**, 380 (1995).

[2]  E. K. Tillinghast, M. A. Townley, T. N. Wight, G. Uhlenbruck, and E. Janssen, Mater. Res. Soc. Symp. Proc. **292**, 9 (1992).

[3]  F. Vollrath, W. J. Fairbrother, R. J. P. Williams, E. K. Tillinghast, D. T. Bernstein, K. S. Gallagher, and M. A. Townley, Nature **345**, 526 (1990).

[4]  F. Vollrath and E. K. Tillinghast, Naturwissenschaften **78**, 557 (1991).

[5]   H. Lee, Nature **465**, 298 (2010).

[6]   V. Sahni, T. A. Blackledge, and A. Dhinojwala, Nat. Commun. **1**, 19 (2010).

[7]   Y. Zhao, M. Morita, and T. Sakamoto, Anal. Sci. **35**, 645 (2019).

[8]   S. Singla, G. Amarpuri, N. Dhopatkar, T. A. Blackledge, and A. Dhinojwala, Nat. Commun. **9**, 1890 (2018).

[9]   G. Amarpuri, C. Zhang, C. Diaz, B. D. Opell, T. A. Blackledge, and A. Dhinojwala, ACS Nano **9**, 11472 (2015).

[10]  V. Sahni, T. A. Blackledge, and A. Dhinojwala, Sci. Rep. **1**, 41 (2011).

[11]  B. D. Opell, D. Jain, A. Dhinojwala, and T. A. Blackledge, J. Exp. Biol. **221**, jeb161539 (2018).

[12]  D. Jain, C. Zhang, L. R. Cool, T. A. Blackledge, C. Wesdemiotis, T. Miyoshi, and A. Dhinojwala, Biomacromolecules **16**, 3373 (2015).

[13]  D. Jain, G. Amarpuri, J. Fitch, T. A. Blackledge, and A. Dhinojwala, Biomacromolecules **19**, 3048 (2018).

[14]  B. D. Opell, C. M. Burba, P. D. Deva, M. H. Y. Kin, M. X. Rivas, H. M. Elmore, and M. L. Hendricks, Ecol. Evol. **9**, 9841 (2019).

[15]  H. Yuk, C. E. Varela, C. S. Nabzdyk, X. Mao, R. F. Padera, E. T. Roche, and X. Zhao, Nature **575**, 169 (2019).

[16]  Y. Zhao, M. Morita, and T. Sakamoto, J. Surf. Anal. **26**, 220 (2019).

[17]  Y. Zhao, M. Morita, and T. Sakamoto, Surf. Interface Anal. **52**, 569 (2020).

[18]  Y. Zhao, M. Morita, and T. Sakamoto, Surf. Interface Anal. (2020). [in press] https://doi.org/10.1002/sia.6924

[19]  N. Scharff and J. A. Coddington, Zool. J. Linn. Soc. **120**, 355 (1997).

[20]  C. E. Griswold, J. A. Coddington, G. Hormiga, and N. Scharff, Zool. J. Linn. Soc. **123**, 1 (1998).

[21]  J. M. Gosline, P. A. Guerette, C. S. Ortlepp, and K. N. Savage, J. Exp. Biol. **202**, 3295 (1999).

[22]  B. O. Swanson, T. A. Blackledge, J. Beltrán, and C. Y. Hayashi, Appl. Phys. A **82**, 213 (2006).

[23]  V. Sahni, D. V Labhasetwar, and A. Dhinojwala, Langmuir **28**, 2206 (2011).

[24]  F. Vollrath and D. T. Edmonds, Nature **340**, 305 (1989).

[25]  T. Köhler and F. Vollrath, J. Exp. Zool. **271**, 1 (1995).

[26]  L. H. Lin, D. T. Edmonds, and F. Vollrath, Nature **373**, 146 (1995).

[27]  H. Elettro, S. Neukirch, F. Vollrath, and A. Antkowiak, Proc. Natl. Acad. Sci. U.S.A. **113**, 6143 (2016).

[28]  M. A. Collin, T. H. Clarke, N. A. Ayoub, and C. Y. Hayashi, Sci. Rep. **6**, 21589 (2016).

[29]  F. Samu and F. Vollrath, Entomol. Exp. Appl. **62**, 117 (1992).

[30]  R. M. Taylor and W. A. Foster, Am. Entomol. **42**, 82 (1996).

[31]  S. Zschokke, Nature **424**, 636 (2003).

[32]  E. Peñalver, D. A. Grimaldi, and X. Delclòs, Science **312**, 1761 (2006).

[33]  E. K. Tillinghast, Naturwissenschaften **68**, 526 (1981).

[34]  O. Choresh, B. Bayarmagnai, and R. V Lewis, Biomacromolecules **10**, 2852 (2009).

[35]  K. Dreesbach, G. Uhlenbruck, and E. K. Tillinghast, Insect Biochem. **13**, 627 (1983).

[36]  E. K. Tillinghast and H. Sinohara, Biochem. Int. **9**, 315 (1984).

[37]  E. K. Tillinghast, Insect Biochem. **14**, 115 (1984).

[38]  H. Schildknecht, P. Kunzelmann, D. Krauss, and C. Kuhn, Naturwissenschaften **59**, 98 (1972).

[39]  E. K. Tillinghast and T. Christenson, J. Arachnol. **12**, 69 (1984).

[40]  M. A. Townley, D. T. Bernstein, K. S. Gallagher, and E. K. Tillinghast, J. Exp. Zool. **259**, 154 (1991).

[41]  G. Amarpuri, V. Chaurasia, D. Jain, T. A. Blackledge, and A. Dhinojwala, Sci. Rep. **5**, 9030

(2015).

[42] V. Sahni, T. Miyoshi, K. Chen, D. Jain, S. J. Blamires, T. A. Blackledge, and A. Dhinojwala, Biomacromolecules **15**, 1225 (2014).

[43] F. Vollrath and D. Edmonds, Naturwissenschaften **100**, 1163 (2013).

[44] G. Amarpuri, C. Zhang, T. A. Blackledge, and A. Dhinojwala, J. R. Soc. Interface **14**, 20170228 (2017).

[45] S. D. Kelly, B. D. Opell, and L. L. Owens, Sci. Nat. **106**, 10 (2019).

[46] B. D. Opell and H. S. Schwend, J. Exp. Zool. **309**, 11 (2008).

[47] E. K. Tillinghast, S. F. Chase, and M. A. Townley, J. Insect Physiol. **30**, 591 (1984).

[48] B. D. Opell, M. E. Clouse, and S. F. Andrews, PLoS One **13**, e0196972 (2018).

[49] F. G. Torres, O. P. Troncoso, and F. Cavalie, Mater. Sci. Eng. C **34**, 341 (2014).

[50] B. D. Opell and M. L. Hendricks, J. Exp. Biol. **213**, 339 (2010).

[51] T. Eisner, R. Alsop, and G. Ettershank, Science **146**, 1058 (1964).

[52] R. E. Peterson and B. J. Tyler, Atmos. Environ. **36**, 6041 (2002).

[53] A. Boulton, D. Fornasiero, and J. Ralston, Int. J. Miner. Process. **70**, 205 (2003).

[54] B. Hagenhoff, D. Breitenstein, E. Tallarek, R. Möllers, E. Niehuis, M. Sperber, B. Goricnik, and J. Wegener, Surf. Interface Anal. **45**, 315 (2013).

[55] T. Stephan, E. K. Jessberger, W. Klöck, H. Rulle, and J. Zehnpfenning, Earth Planet. Sci. Lett. **128**, 453 (1994).

[56] T. Stephan, A. L. Butterworth, F. Hörz, C. J. Snead, and A. J. Westphal, Meteorit. Planet. Sci. **41**, 211 (2006).

[57] J. Leitner, T. Stephan, A. T. Kearsley, F. Hörz, G. J. Flynn, and S. A. Sandford, Meteorit. Planet. Sci. **43**, 161 (2008).

[58] T. Stephan, D. Rost, E. P. Vicenzi, E. S. Bullock, G. J. MacPherson, A. J. Westphal, C. J. Snead, G. J. Flynn, S. A. Sandford, and M. E. Zolensky, Meteorit. Planet. Sci. **43**, 233 (2008).

[59] T. Stephan, G. J. Flynn, S. A. Sandford, and M. E. Zolensky, Meteorit. Planet. Sci. **43**, 285 (2008).

[60] K. G. Stowe, S. L. Chryssoulis, and J. Y. Kim, Miner. Eng. **8**, 421 (1995).

[61] C. Piantadosi, M. Jasieniak, W. M. Skinner, and R. S. C. Smart, Miner. Eng. **13**, 1377 (2000).

[62] C. Piantadosi and R. S. C. Smart, Int. J. Miner. Process. **64**, 43 (2002).

[63] T. Sakamoto, M. Koizumi, J. Kawasaki, and J. Yamaguchi, Appl. Surf. Sci. **255**, 1617 (2008).

[64] R. E. Peterson and B. J. Tyler, Appl. Surf. Sci. **203**, 751 (2003).

[65] B. Tomiyasu, K. Suzuki, T. Gotoh, M. Owari, and Y. Nihei, Appl. Surf. Sci. **231**, 515 (2004).

[66] S. Sobanska, G. Falgayrac, J. Rimetz-Planchon, E. Perdrix, C. Bremard, and J. Barbillat, Microchem. J. **114**, 89 (2014).

[67] D. J. Gaspar, A. Laskin, W. Wang, S. W. Hunt, and B. J. Finlayson-Pitts, Appl. Surf. Sci. **231**, 520 (2004).

[68] G. S. Groenewold, J. C. Ingram, T. McLing, A. K. Gianotto, and R. Avci, Anal. Chem. **70**, 534 (1998).

[69] Y.-J. Zhu, N. Olson, and T. P. Beebe, Environ. Sci. Technol. **35**, 3113 (2001).

[70] J. Coumbaros, K. P. Kirkbride, G. Klass, and W. Skinner, Forensic Sci. Int. **119**, 72 (2001).

[71] J. Dubochet and A. W. McDowall, J. Microsc. **124**, 3 (1981).

[72] N. Becker, E. Oroudjev, S. Mutz, J. P. Cleveland, P. K. Hansma, C. Y. Hayashi, D. E. Makarov, and H. G. Hansma, Nat. Mater. **2**, 278 (2003).

[73] C. Diaz, D. Maksuta, G. Amarpuri, A. Tanikawa, T. Miyashita, A. Dhinojwala, and T. A. Blackledge, J. R. Soc. Interface **17**, 20190792 (2020).

[74] J. M. Gosline, M. E. DeMont, and M. W. Denny,

Endeavour **10**, 37 (1986).

[75] C. M. Anderson and E. K. Tillinghast, Physiol. Entomol. **5**, 101 (1980).

[76] K. Kanenari, M. Morita, and T. Sakamoto, E-J. Surf. Sci. Nanotechnol. **14**, 131 (2016).

[77] A. L. F. de Barros, E. F. da Silveira, L. S. Farenzena, and K. Wien, J. Surf. Investig. **7**, 1225 (2013).

[78] R. Souda, Phys. Rev. Lett. **93**, 235502 (2004).

[79] M. A. Robinson and D. G. Castner, Biointerphases **8**, 15 (2013).

[80] R. Souda, Phys. Rev. B **70**, 165412 (2004).

[81] R. Souda, J. Chem. Phys. **121**, 8676 (2004).

[82] R. Martinez, A. N. Agnihotri, P. Boduch, A. Domaracka, D. Fulvio, G. Muniz, M. E. Palumbo, H. Rothard, and G. Strazzulla, J. Phys. Chem. A **123**, 8001 (2019).

[83] R. Van Ham, A. Adriaens, L. Van Vaeck, R. Gijbels, and F. Adams, Nucl. Instrum. Methods Phys. Res. **161**, 245 (2000).

[84] Y. R. Shen and V. Ostroverkhov, Chem. Rev. **106**, 1140 (2006).

[85] C. Diaz, A. Tanikawa, T. Miyashita, G. Amarpuri, D. Jain, A. Dhinojwala, and T. A. Blackledge, R. Soc. Open Sci. **5**, 181296 (2018).